\documentclass[a4paper,usenatbib]{mnras}
\usepackage{newtxtext,newtxmath}
\usepackage[T1]{fontenc}
\usepackage{ae,aecompl}

\usepackage{graphicx}
\usepackage{amsmath}
\usepackage{amssymb}
\newcommand{\cp}{\citep}
\newcommand{\ct}{\citet}

\title[Predictions for the Transition Between Rocky Super-Earths and Gaseous Sub-Neptunes]{How Formation Timescales Affect the Period Dependence of the Transition Between Rocky Super-Earths and Gaseous Sub-Neptunes and Implications for $\eta_{\mathrm{\oplus}}$}
\author[Eric D. Lopez, Ken Rice]{Eric D. Lopez,$^{1,2,3}$\thanks{E-mail: elopez@roe.ac.uk} \& Ken Rice$^{3,4}$
\\
$^{1}$NASA Goddard Space Flight Center, 8800 Greenbelt Rd, Greenbelt, MD, 20771, USA\\
$^{2}$GSFC Sellers Exoplanet Environments Collaboration, NASA GSFC, Greenbelt, MD 20771\\
$^{3}$SUPA, Institute for Astronomy, Royal Observatory Edinburgh, University of Edinburgh, Blackford Hill, Edinburgh, UK, EH9 3HJ\\
$^{4}$ Centre for Exoplanet Science, University of Edinburgh, Edinburgh, UK\\
}

\pubyear{2016}

\begin{document}
\label{firstpage}
\pagerange{\pageref{firstpage}--\pageref{lastpage}}
\maketitle

\begin{abstract}

One of the most significant advances by NASA's {\it Kepler} Mission was the discovery of an abundant new population of highly irradiated planets with sizes between those of the Earth and Neptune, unlike anything found in the Solar System. Subsequent analysis showed that at $\sim$1.5 $R_{\mathrm{\oplus}}$ there is a transition from a population of predominantly rocky super-Earths to non-rocky sub-Neptunes, which must have substantial volatile envelopes to explain their low densities. Determining the origin of these highly irradiated rocky planets will be critical to our understanding of low-mass planet formation and the frequency of potentially habitable Earth-like planets. These short-period rocky super-Earths could simply be the stripped cores of sub-Neptunes, which have lost their envelopes due to atmospheric photo-evaporation or other processes, or they might instead be a separate population of inherently rocky planets, which never had significant envelopes. We suggest an observational path forward to distinguish between these scenarios. Using models of atmospheric photo-evaporation, we show that if most bare rocky planets are the evaporated cores of sub-Neptunes then the transition radius should {\it decrease} as surveys push to longer orbital periods, since on wider orbits only planets with smaller less massive cores can be stripped. On the other hand, if most rocky planets formed after their disks dissipate then these planets will have formed without initial gaseous envelopes. In this case, we use N-body simulations of planet formation to show that the transition radius should {\it increase} with orbital period, due to the increasing solid mass available in their disks. Moreover, we show that distinguishing between these two scenarios should be possible in coming years with radial velocity follow-up of planets found by TESS. Finally, we discuss the broader implications of this work for current efforts to measure $\eta_{\mathrm{\oplus}}$, which may yield significant overestimates if most rocky planets form as evaporated cores.

\end{abstract}

\begin{keywords}
planets and satellites: atmospheres, planets and satellites: composition, planets and satellites: physical evolution
\end{keywords}

\section{Introduction}

One of the key revelations from NASA's Kepler Mission has been the discovery of an abundant new population of short period planets with transit radii in between the radii of Earth and Neptune. These planets occupy a range of sizes and orbits that is completely vacant in the Solar System, and so they present a key test of traditional models of planet formation. Below a transit radius of $\sim$1.5 $R_{\mathrm{\oplus}}$, most planets with measured masses seem to be consistent with bare rocky compositions with an Earth-like mixture of silicates and iron \cp{Dressing2015}. However, at $\sim$1.5 $R_{\mathrm{\oplus}}$ there appears to be a transition between a primarily rocky and a primarily non-rocky planet population, where most planets above this size must have large volatile envelopes to explain their lower densities \cp{Rogers2015}. Throughout this paper, we will refer to the radius at which this transition between these rocky super-Earths and non-rocky sub-Neptunes as $R_{\mathrm{trans}}$. 

One key question is whether the rocky super-Earths below $R_{\mathrm{trans}}$ and the volatile rich sub-Neptunes above $R_{\mathrm{trans}}$ represent a single continuous population whose initial compositions have been sculpted by post-formation evolution, or two separate populations with distinct formation mechanisms. In the first case, both the rocky super-Earths and non-rocky sub-Neptunes would form quickly, while they were still embedded in their gaseous proto-stellar disks, which typically last up to $\sim$10 Myr \cp{Haisch2001, Williams2011}. The planets would quickly form their rocky or icy cores and would then accrete gaseous envelopes directly from the disk \cp[e.g.,][]{Bodenheimer1986, Lissauer1993, Pollack1996}. If planets reach $\sim$half of their total mass in gas, then they will typically undergo run-away accretion to produce gas giants \cp[e.g.,][]{Pollack1996, Alibert2005, Rafikov2006}. However, most short-period planets should remain well below this limit and we would instead expect them to have a broad range of initial envelope mass fractions from <1\% gas to $\sim$50\% by mass as determined primarily by disk mass and lifetime, the opacity of dust grains, planetary core mass, the local disk temperature, and orbit of the planets \cp[e.g.,][]{Rogers2011, Ikoma2012, Mordasini2012a, Bodenheimer2014, Mordasini2014, Lee2015, Lee2016}. The short-period rocky planets meanwhile, would represent those planets which lost their initial gaseous envelopes through processes like XUV-driven photo-evaporation \cp[e.g.,][]{Lopez2012,Owen2013,Jin2014,Chen2016,Lopez2016} or atmospheric impact erosion \cp[e.g.,][]{Catling2013, Inamdar2015, Schlicting2015, Liu2015, Inamdar2016}. 

On the other hand, if the rocky planet population finished assembling their cores after their proto-stellar disks had already dissipated, then they would never have had initial gaseous envelopes and would represent a primordial rocky planet population. This is generally believed to be how the terrestrial planets of the inner solar system formed, where evidence from isotope ratios \cp[e.g.,][]{Tera1974,Allegre1995, Touboul2007, Kleine2009, Nemchin2009} suggest that the Earth mostly finished assembling with the moon forming impact when the solar system was $\sim$30-100 Myr old. While there is a great deal of debate about the exact timing \cp[e.g., see][]{Kleine2009}, the final phases of Earth's formation very likely took place long after the proto-solar nebula had already dissipated \cp{Lissauer1987, Haisch2001}. This is not to say that a primordial rocky planet population would consist of completely airless bodies, they could have significant secondary atmospheres reaching pressures of hundred of bars that are outgassed from their interiors \cp{Elkins-Tanton2008, Schaefer2010}. However, any such secondary atmospheres would still only represent a small fraction of a planets total mass $\lesssim$0.1\% Although extremely important for habitability,  a hydrogen envelope that is $<10^{-3}$ $M_{\mathrm{\oplus}}$ would be far too thin to measurably change the bulk radius of a transiting planet \cp{Lopez2014}, and so from the point of view of current transit surveys any planets without large gaseous envelopes accreted from the disk, including the Earth, are indistinguishable from bare rocks.

\begin{figure}
  \begin{center}
    \includegraphics[width=3.5in,height=2.5in]{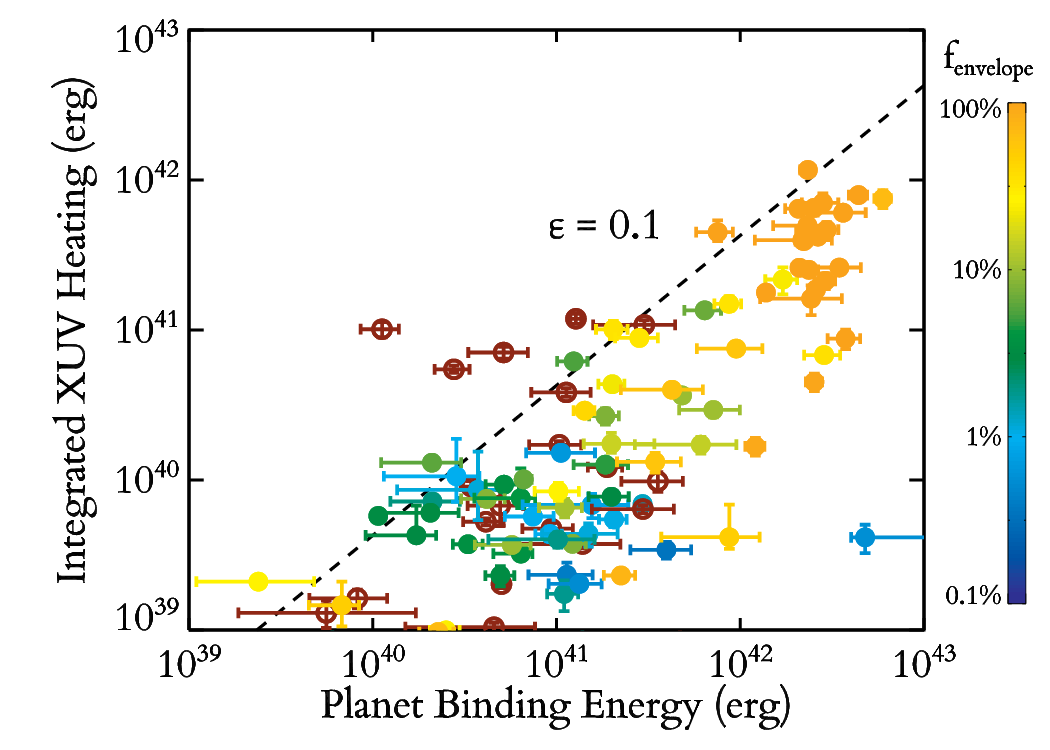}
  \end{center}
  \caption{Updated from \ct{Lopez2014} this shows the total lifetime XUV heating planets receive, assuming their current orbits and radii, vs. their current gravitational binding energy. Included are all currently known transiting planets below 100 $M_{\mathrm{\oplus}}$ with masses and radii measured to better than 50\%. Planets are color-coded by their H/He envelope mass fractions, assuming Earth-like cores, with planets that are likely bare rock shown by by rust-colored open circles. The dashed line meanwhile corresponds to the photo-evaporation threshold predicted by \ct{Lopez2012}, above which we find only bare rocky planets, which have likely been stripped of any primordial envelope. \label{bindfig}}
\end{figure}

Both these scenarios, a primordial rocky planet population and one that originated as the stripped cores of gaseous sub-Neptunes, are consistent with current evidence. This is because the vast majority of confirmed rocky exoplanets are on highly irradiated orbits where planets are extremely vulnerable to losing any primordial envelopes to atmospheric escape \cp[e.g.,][]{Leger2009, Batalha2011, Pepe2013, Dressing2015}. Indeed, from the overall distribution of short-period low-mass transiting planets, there is substantial evidence that this population has been significantly sculpted by XUV-driven photo-evaporation, or another comparable process \cp[e.g.,][]{Lopez2012, Jackson2012,Owen2013,Jin2014,Chen2016}. For example, in Figure \ref{bindfig} we compare the current gravitational binding energy of all low-mass transiting planets to the lifetime integrated X-ray heating these planets have received. This shows a clear threshold, consistent with models of photo-evaporation, beyond which no known planets have retained gaseous H/He envelopes \cp{Lopez2012,Lopez2014,Owen2012,Owen2013,Jackson2012,Chen2016,Jin2014}. Moreover, starting from a log-uniform initial distribution of envelope mass fractions \ct{Chen2016} recently showed that photo-evaporation models can adequately reproduce the overall distribution of planet radii for planets with orbits out to $\sim$0.25 AU.

Critically, however, because of its relatively low XUV irradiation, the Earth could not have formed as the evaporated core of a sub-Neptune. Although very early studies by \ct{Hayashi1979} and \ct{Sekiya1980} suggested that Earth could have had a significant initial hydrogen envelope, these studies were conducted before we had good observational constraints on disk lifetimes and stellar XUV histories. Subsequent studies by \ct{Erkaev2013} and \ct{Lammer2014} estimate that the Earth could have only lost up to $\sim$10 Earth Ocean equivalents of hydrogen, which corresponds to $\sim3\times10^{-4}$ $M_{\mathrm{\oplus}}$ or a surface pressure of $\sim$50 bar. A more recent study by \ct{Johnstone2015} found that the Earth could have lost up to 1\% of it's mass, however this required assuming both that the Sun was born as a very fast rotator and an extremely high core luminosity. Consequently, $10^{-3}$ $M_{\mathrm{\oplus}}$ is likely a reasonable estimate for the maximum primordial envelope that can be evaporated from a typical Earth-like planet, and therefore we consider Earth to have been ``born-rocky". 

As a result, distinguishing between these two scenarios for the origin of the rocky exoplanet population will be critical in our efforts to constrain the frequency of Earth-like planets in the habitable zones of Sun-like stars, as discussed in Section \ref{etasec}. To do this, we propose a new observational test to constrain the origin of the rocky planet population by determining how the transition radius $R_{\mathrm{trans}}$ between the rocky and non-rocky planet populations scales with the incident flux a planet receives from its star. Using N-body simulations of planet assembly and models of atmospheric photo-evaporation we show that these two scenarios make opposing predictions for how the transition should scale with flux, thereby providing us with a powerful diagnostic for the formation of rocky planets.

\section{Scenario 1: Rocky Planets as the Stripped Cores of Hot Neptunes}\label{evapsec}
\subsection{Evaporation Model}
In order to make predictions for the impact of photo-evaporation on the rocky to non-rocky transition we use of the planet evolution model described in \ct{Lopez2016}. \ct{Lopez2014} and \ct{Lopez2016} describe this model in greater detail, however, for the benefit of the reader we briefly summarize the key features here. The evolution model consists of two main components. The first is an interior structure and thermal evolution model, which computes hydrostatic structure models for planets with gaseous envelopes atop rocky cores and then evolves them in time as a planet cools and contracts after formation \cp{Lopez2012,Lopez2014}. This allows us to predict planetary radii as a function of a planet's core mass, envelope mass, irradiation, and age. The models presented here are computed for solar-composition H/He envelopes atop Earth-like rocky cores with 1/3 of their mass in and iron core and 2/3 in a silicate mantle orbiting Sun-like stars. We have made one small modification to the structure model from previous publications, to maintain consistency with the later discussion we now use the core mass-radius relationship from \ct{Zeng2016}. This thermal evolution model is then coupled to a parameterized photo-evaporation model as described in \ct{Lopez2012} and \ct{Lopez2013}, which allows us to predict how the envelope mass and therefore radius change as a planet loses mass to XUV-driven photo-evaporation. Here we use the modified mass-loss prescription described in \ct{Lopez2016}, which accounts for the impact of radiation-recombination limited cooling  \cp{Murray-Clay2009} for the most extremely irradiated planets. Specifically, we compute the overall mass loss rate by taking the minimum of the energy-limited mass loss rate using the formalism of \ct{Erkaev2007}, which is applicable for planets receiving XUV fluxes higher than Earth today, and the radiation-recombination limited rate, which is appropriate for more irradiated planets, following the formalism of \ct{Murray-Clay2009} and \ct{Chen2016}. These two rates are described by equations (\ref{eleq}) and (\ref{rrtotaleq})

\begin{equation}\label{eleq}
\dot{M}_{\mathrm{EL}}  = -\frac{\epsilon_{\mathrm{XUV}}  \pi F_{\mathrm{XUV}} R_{\mathrm{base}}^3 }{G M_{\mathrm{p}} K_{\mathrm{tide}} }
\end{equation}

\begin{equation}\label{rrtotaleq}
\begin{aligned}
\dot{M}_{\mathrm{RR}}  = {} &-4 \pi c_{\mathrm{s}} R_{\mathrm{s}}^2  \mu_{\mathrm{+,wind}} m_{\mathrm{H}} \left( \frac{F_{\mathrm{XUV}} G M_{\mathrm{p}}  }{h \nu_{ \mathrm{0}} \alpha_{\mathrm{rec,B}}  c_{\mathrm{s}}^2 R_{\mathrm{base}}^2 }  \right)^{1/2} \\
& \times \, \exp{\left[ \frac{ G M_{\mathrm{p}} }{ R_{\mathrm{base}} c_{\mathrm{s}}^2 } \left( \frac{ R_{\mathrm{base}} }{ R_{\mathrm{s}} } - 1 \right)  \right]}
\end{aligned}
\end{equation}


Here $\epsilon_{\mathrm{XUV}}$ is a parameterization of the efficiency of photo-evaporation, generally taken to be $\sim$10\% for solar composition atmospheres \cp[e.g.,][]{Jackson2010, Valencia2010, Lopez2012, Jin2014, Chen2016}. $F_{\mathrm{XUV}}$ is the XUV flux at a planet's orbit, which we take as $F_{\mathrm{XUV}} = 29.7 (L_{\mathrm{s}}/L_{\mathrm{\odot}})(a/\mathrm{AU})^{-2}(age/\mathrm{Gyr})^{-1.23}$ erg s$^{-1}$ cm$^{-2}$ for Sun-like stars older than 100 Myr based on \ct{Ribas2005}, while at ages younger than 100 Myr we assume that the stellar XUV luminosity saturates at $\approx 10^(-3.5)L_{\mathrm{bol}}$ based on \ct{Jackson2012}. $R_{\mathrm{base}}$ and $R_{\mathrm{s}}$ are the radii of the XUV photosphere and the sonic point respectively, computed following the method described in \ct{Lopez2016}, and $c_{\mathrm{s}}$ is the sound speed at the sonic point, typically $\sim$10 km/s. $M_{\mathrm{p}}$ is the total planet mass. $K_{\mathrm{tide}}$ is a slight geometric correction factor. Finally, $h\nu_{ \mathrm{0}}\approx20$ eV is the typically energy of the incoming ionizing radiation and $\alpha_{\mathrm{rec,B}}$ is the case B recombination coefficient for hydrogen.

Taking the minimum of these two rates is a commonly used approximation \cp[e.g.,][]{Jin2014,Chen2016}, which approximates the predictions of hydrodynamic mass loss models \cp[e.g.,][]{Murray-Clay2009, Owen2012}, and is generally applicable for planets with H/He envelopes and periods $\lesssim$100 days, where the evaporative wind should be fully collisional. For planets in the habitable zones of Sun-like stars this model is not applicable, since there it is necessary to take into account the role of molecular coolants and conduction \cp[e.g.,][]{Tian2008}, along with non-collisional and non-thermal escape processes, however, modeling such planets is beyond the scope of this paper and at any rate they are not relevant to the observational predictions made here.

\subsection{Evaporation Results}

Using this model, we then ran a large suite of approximately 20,000 evolution models on a grid covering a range of initial core masses, envelope fractions, and levels of irradiation. The points on this grid were spaced uniformly in log space with cores ranging from 1 to 20 $M_{\mathrm{\oplus}}$, initial envelope fractions from 0.1 to 50\%, and bolometric incident fluxes from 10 to 1000 $F_{\mathrm{\oplus}}$. We chose this log-uniform spacing primarily to fully explore the relevant parameter space. However, as we noted before, \ct{Chen2016} found that such an initial distribution was able to reproduce the observed radius distribution when photo-evaporation is included. In any case, the general predictions for the flux dependence of the transition radius presented here are insensitive to any of these choices. We allowed these models to start photo-evaporating at 10 Myr, shortly after the end of planet formation, and ended them once the planet reached 5 Gyr, at which point we recorded the final planet radius and envelope fraction.

\begin{figure}
  \begin{center}
    \includegraphics[width=3.4in,height=2.43in]{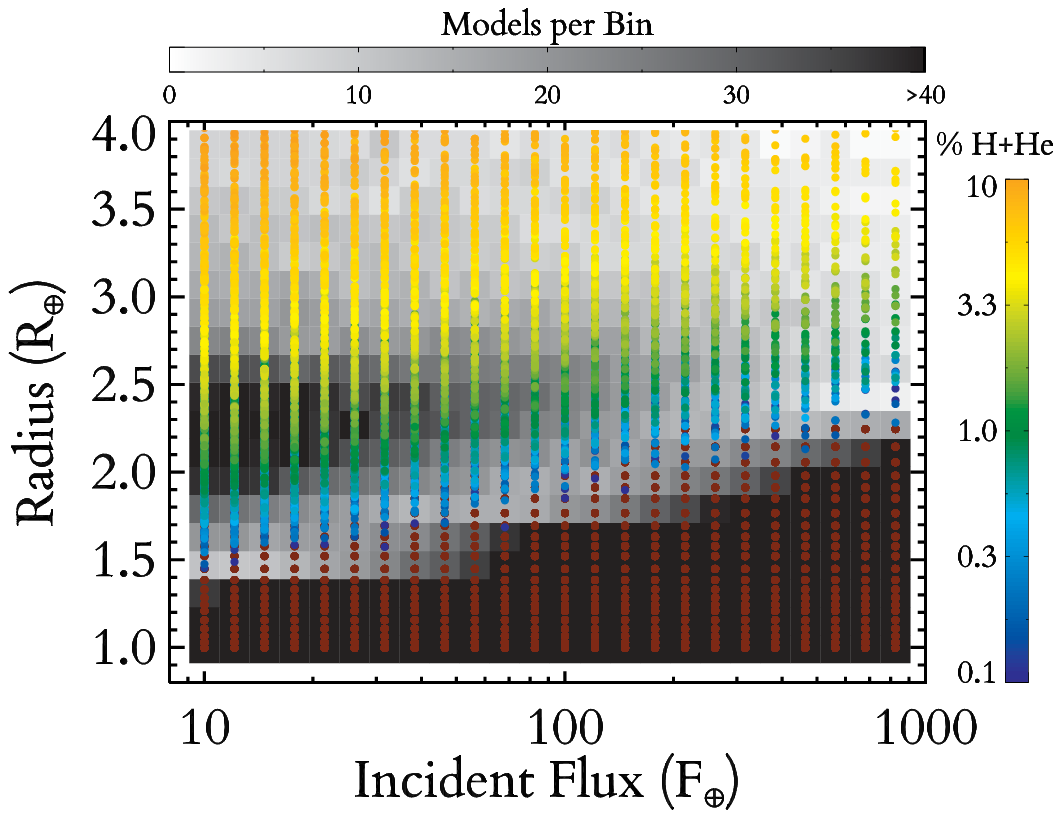}
  \end{center}
  \caption{This plots the final planet radius predicted by the evolution model after 5 Gyr of thermal and photo-evaporative evolution vs. the incident bolometric flux that a planet receives at its orbit, for planets with solar composition H/He envelopes atop Earth-like cores. $\gtrsim$ 20,000 individual model runs where performed to generate this figure. The results of individual runs are shown by the points, which have been color-coded by their final H/He envelope mass fraction. Rust-colored points in the bottom right indicate bare rocky planets which have completely lost their H/He envelopes. The grey-scale background meanwhile shows the number of models that ended up in each radius-flux bin, where darker shades corresponds to a higher density of points, and clear regions correspond to areas devoid of models. \label{valleyfig}}
\end{figure}

\begin{figure}
  \begin{center}
    \includegraphics[width=3.4in,height=2.43in]{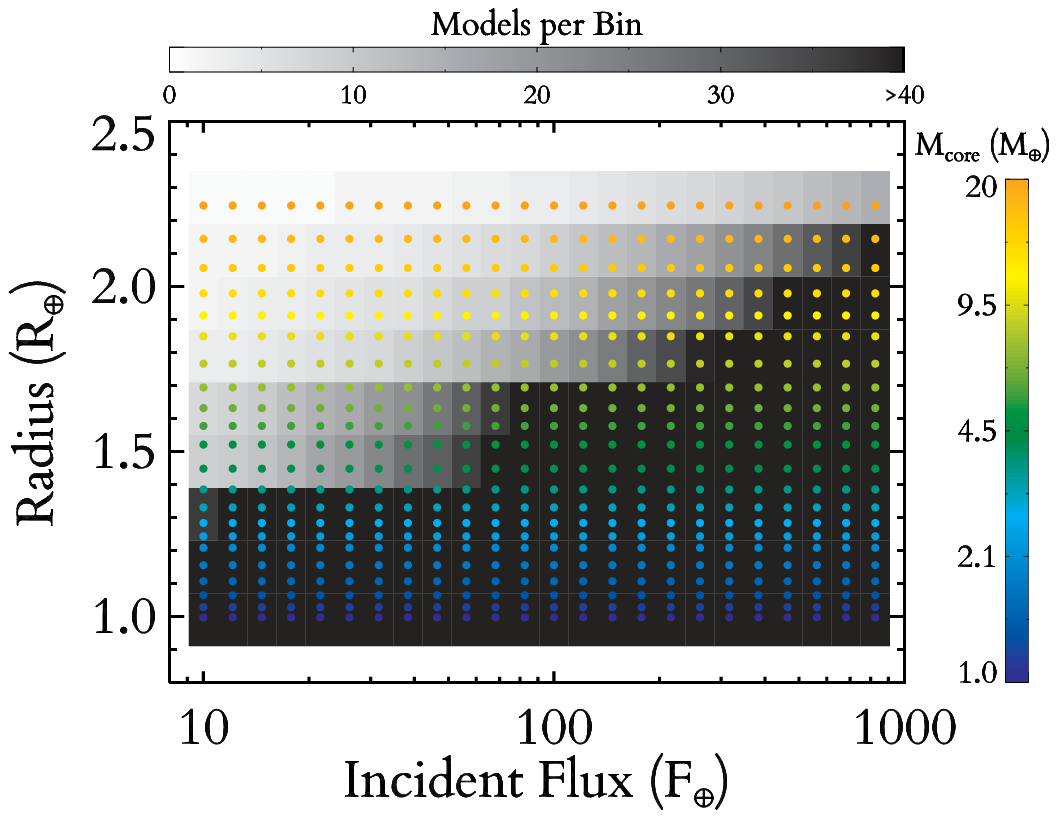}
  \end{center}
  \caption{Similar to Figure \ref{valleyfig} except here we only show those rocky planets that have completely lost their gaseous envelopes. Again, the grey-scale shows the number of models that ended up in each radius-flux bin, where darker shades corresponds to a higher density of models. The color-coding meanwhile shows the core-mass for each of our models. Critically, if most rocky planets originate as the evaporated remnants of sub-Neptunes then the maximum size of bare-rocky planets should {\it increase} with increasing incident flux (decreasing orbital period).  \label{evapfig} }
\end{figure}

Figure \ref{valleyfig} summarizes the results from this grid of models. At lower levels of irradiation $\lesssim$100 $F_{\mathrm{\oplus}}$ and larger radii $\gtrsim$ 1.5$R_{\mathrm{\oplus}}$ we find the population of gas rich sub-Neptunes which have resisted photo-evaporation. These are the most abundant population of exoplanets found by {\it Kepler} \cp{Petigura2013a,Burke2015} and our model predicts that planets in this size range typically have moderate gas envelopes composing $\sim$1-10\% of their total mass \cp{Lopez2014}. Meanwhile, at higher levels of irradiation and smaller sizes we find the population of bare rocky cores that have had their envelopes completely stripped away by photo-evaporation. These are the planets that we are interested in here, and we will discuss the features of this population more below. Finally, in between these two populations there is a narrow ``evaporation valley" in which our model and other evaporation models \cp[e.g.,][]{Owen2013,Jin2014}, predict that planets should be relatively rare. Although it is not the focus of this paper, the evaporation valley is also a key prediction of these models, which may be diagnostic in constraining the formation and composition of sub-Neptunes and super-Earths, particularly whether they contain large amounts of water or other volatiles ices that formed beyond the snow-line \cp{Lopez2013,Owen2013}. Indeed, using updated stellar parameters from the CKS survey \cp{Petigura2017,Johnson2017} of {\it Kepler} host stars \ct{Fulton2017} recently found convincing evidence for the existence of this valley or gap in the radius distribution of {\it Kepler} planets with orbital periods $\lesssim$50 days around Sun-like stars. Morever, \ct{Owen2017} and \ct{Jin2017} subsequently showed that the observed valley seen by \ct{Fulton2017} is well reproduced by photo-evaporation similar to those used here.

In Figure \ref{evapfig} we have isolated the sample of stripped rocky cores in order to focus in on the features of this population. Here we can clearly see that the maximum size of stripped rocky planets produced by photo-evaporation increases with the incident flux a planet receives, and therefore decreases with increasing orbital period. This is because planets with more massive rocky cores are more resistant to photo-evaporation and therefore require greater irradiation to lose their envelopes. \ct{Lopez2013} showed that the incident flux necessary for a planet to lose its envelope scales roughly as $F_{\mathrm{p}} \propto M_{\mathrm{core}}^{2.4}$. Therefore at higher fluxes planets with more massive cores can be stripped, and these more massive cores are correspondingly larger. At fixed composition, the radius of Earth-like rocky planets scale roughly as $R_{\mathrm{p}} \propto M_{\mathrm{p}}^{1/3.7}$ \cp{Zeng2016}. Combining these two factors then, we would expect that if the rocky planet population is primarily produced by photo-evaporative stripping of sub-Neptunes then the transition radius between rocky and non-rocky planets should scale roughly as $R_{\mathrm{trans}} \propto F_{\mathrm{p}}^{0.11}$, which is consistent with what we see in Figures \ref{valleyfig} and \ref{evapfig}. 

Note here that in this scenario what matters is the total lifetime XUV heating that a planet receives $F_{\mathrm{XUV, lifetime}}$. At a given stellar type, the average XUV heating will be proportional to a planet's current bolometric incident flux. However, at later stellar types $F_{\mathrm{XUV}}$ increases relative to $F_{\mathrm{bol}}$, so this must be taken into account when comparing planets across a range of host stellar types. Indeed, using X-ray observations of FGK stars from \ct{Jackson2012} and FUV observations of early to mid M dwarfs from \ct{Shkolnik2014}, McDonald et al. (submitted to ApJ) recently calculated the expected lifetime X-ray heating received by {\it Kepler} planets and found that at fixed present day bolometric flux this scales roughly as $E_{\mathrm{XUV, lifetime}}/F_{\mathrm{bol, current}}\propto M_{\mathrm{s}}^{-3}$. X-ray, FUV, and XUV flux should generally be closely correlated, and so a rocky/non-rocky transition produced by photo-evaporation should produce a transition radius that scales as 

\begin{equation}\label{evaptranseq}
R_{\mathrm{trans,evap}} \propto F_{\mathrm{p}}^{0.11} M_{\mathrm{s}}^{-0.33}.
\end{equation}

Alternatively, we can rewrite this in time of orbital period since the incident bolometric flux a planet receives is simply $F_{\mathrm{p}}=L_{\mathrm{s}}/(4\pi a^2)$, where $Ls \propto Ms^4$ for FGK and early M stars with $0.43\leq M_{\mathrm{s}} \leq2.0$ \cp{Kippenhahn1990}. Therefore using Kepler's law, $F_{\mathrm{p}} \propto L_{\mathrm{s}} M_{\mathrm{s}}^{-2/3} P^{-4/3} \propto M_{\mathrm{s}}^{10/3} P^{-4/3}$. Putting this in equation (\ref{evaptranseq}), then we find that 

\begin{equation}\label{evaptranseq2}
R_{\mathrm{trans,evap}} \propto P^{-0.15} M_{\mathrm{s}}^{0.04}.
\end{equation}

Although, given the observational uncertainties on $F_{\mathrm{XUV, lifetime}}/F_{\mathrm{bol, current}}$, this is also consistent with having no stellar mass dependence when the transition radius is written relative to period, so $R_{\mathrm{trans,evap}} \propto P^{-0.15}$.

One key caveat with these results is that we have used just a single evolution track for the stellar XUV age relation for Sun-like stars. However, recent studies \cp[e.g.,][]{Tu2015} have shown that among Sun-like stars the evolution of stellar XUV luminosity can depend strongly on a star's initial rotation rate, with stars with faster initial rotation staying in the saturation regime for a much larger fraction of their lifetimes. Comparing stellar evolution tracks for G stars at the 10th and 90th percentiles in initial rotation, we estimate that differences in initial rotation could lead to up to factor of $\approx7\times$ in the total lifetime integrated XUV luminosity from a Sun-like star at an age of 5 Gyr. Consequently, the observed spread in stellar rotation rates may translate into spread in the location of the rocky/non-rocky transition, with the largest stripped rocky cores likely occurring around the fastest rotating stars.

Fortunately, however, the slopes of the scaling relations derived above are independent of this scatter in initial rotation rates since these are set by the flux and mass dependence of the photo-evaporative escape process together with the rocky core mass distribution. Therefore so long as there is no correlation between stellar rotation rates and the core masses of low-mass planets then our results should be unchanged. In short, changes in stellar rotation and XUV emission evolution will affect the location of the evaporation valley, but not it's slope. This does however, suggest that stellar rotation history may be a valuable extra dimension when examining the trends in the observed {\it Kepler} population, something which should be increasingly possible with dramatically improved stellar parameters thanks to spectroscopic follow-up surveys \cp[e.g.,][]{Petigura2017,Johnson2017}, asteroseismology \cp[e.g.,][]{Silva2015,VanEylen2018}, and Gaia parallaxes \cp[e.g.,][]{Berger2018}.

\subsection{Additional Caveats: Impact Erosion and Water Worlds}

Before we move on to our second scenario, there two other possible caveats that we wish to address here. First, as previously mentioned, some studies \cp[e.g.,][]{Catling2013, Inamdar2015, Schlicting2015, Liu2015, Inamdar2016} have suggested that atmospheric erosion by impacts may play a large role in sculpting the compositions of short-period planets in a manner comparable to mass-loss due to photo-evaporation. However, a population which has been stripped by impact erosion  will have a different correlation between planet mass, orbital period, and stellar mass than one which has been sculpted by photo-evaporation. For example, in the limit of small impactors, \ct{Schlicting2015} found that to first order the final mass loss fraction went as $X_{\mathrm{loss}} \propto (m_{\mathrm{imp}}v_{\mathrm{imp}})/(M_{\mathrm{p}}v_{\mathrm{esc}})$, where $m_{\mathrm{imp}}$ and $v_{\mathrm{imp}}$ are the mass and velocity of the impactor and and $v_{\mathrm{esc}} = \sqrt{2GM_{\mathrm{p}}/R_{\mathrm{p}}}$ is the planet's escape velocity. Assuming circular orbits and that $v_{\mathrm{imp}} \propto v_{\mathrm{orb}}=2\pi a/P$, then we would expect that $M_{\mathrm{p}}^{3/2}R_{\mathrm{p}}^{-1/2} \propto M_{\mathrm{s}}^{1/3} P^{-1/3}$. Again using $M_{\mathrm{p}} \propto R_{\mathrm{p}}^{3.7}$ for rocky planets from \ct{Zeng2016}, then this approximately gives $M_{\mathrm{p}}^{5} \propto (M_{\mathrm{s}}/P)^{1/3}$ or 

\begin{equation}
R_{\mathrm{trans,impacts}} \propto (M_{\mathrm{s}}/P)^{1/15}.  
\end{equation}

Again using $F_{\mathrm{p}} \propto M_{\mathrm{s}}^{10/3} P^{-4/3}$ this can then be re-written as

\begin{equation}
R_{\mathrm{trans,impacts}} \propto F_{\mathrm{p}}^{1/20}  M_{\mathrm{s}}^{-1/6}.
\end{equation}

This is only an approximation and it neglects second order terms that \ct{Schlicting2015} found were important for large impactors. Nonetheless, this clearly predicts that the variation of the transition radius with irradiation and stellar type for planets that have been stripped by impacts should be weaker than what we found above for photo-evaporation in equations (\ref{evaptranseq}) and (\ref{evaptranseq2}). 

Finally, in addition to bare rocky planets with Earth-like compositions and planets with modest H/He envelopes atop Earth-like cores, it is also possible that, like Uranus and Neptune, some short-period exoplanets could accrete much of their mass from water and other volatile ices if they or their planetesimal building blocks migrated from beyond the snow-line \cp[e.g.,][]{Rogers2011, Hansen2012, Mordasini2012a}. For individual planets, it is not possible to rule out this possibility using planetary mass and radius alone \cp{Rogers2010a}. However, water-dominated envelopes should be much more resilient against photo-evaporation than solar composition envelopes, and so there should be very few planets that are the stripped cores of former water-worlds and those should only be found on the most extreme ultra-short-period orbits \cp{Lopez2016}.

\section{Scenario 2: A Primordial Rocky Population Born after Disk Dispersal}\label{rockysec}

The second possibility is that the rocky planet population simply never had significant volatile envelopes in the first place and that they formed with their current essentially bare rocky compositions. This would make sense if these planets took $\gtrsim$10 Myr to finish assembling, as by that point their proto-planetary gas disks will have already dissipated. As mentioned above, this is generally believed to be how the Earth and other terrestrial planets of the inner Solar System finished forming \cp[e.g.,][]{Raymond2009,Morbidelli2012}.

In this case the rocky and non-rocky exoplanets would represent two separate populations, likely originating from different formation timescales, and the transition radius $R_{\mathrm{trans}}$ rocky and non-rocky planet would instead arise from the superposition of these two populations.  Therefore in this scenario, the transition radius will be set by the maximum typical mass of bare rocky planets as a function of distance from the star, which should simply be set by the available supply of solid materials that a planetary core can accrete by collisions. This is in turn will beset by the solid density profile in an average disk. To get an idea for the typical solid density profile from which exoplanetary systems formed we turn to the analysis of \ct{Chiang2013}. By estimating the heavy element masses for the full sample of short-period planets found by {\it Kepler} as a function of their semi-major axes, \ct{Chiang2013} were able to construct a typical Minimum Mass Extrasolar Nebula (MMEN). Specifically they found that the {\it Kepler} sample implies a typical initial solid surface density profile of the form

\begin{equation}\label{mmeneq}
\sigma_{\mathrm{solid}}=6.2\times10^2 \mathcal{F}_{\mathrm{disc}} (a/0.2 \, \mathrm{AU})^{-1.6} \, \mathrm{g \, cm^{-2}}.
\end{equation}

Here, $a$ is semi-major axis and $\mathcal{F}_{\mathrm{disc}}$ is a normalization factor that can vary the surface density relative to the MMEN. This is quite similar in form to the standard overall surface density profile for the Minimum Mass Solar Nebula $\sigma(a)=\sigma_{0}(a/\mathrm{AU})^{-3/2}$ \cp{Hayashi1981}, except that the MMEN requires $\sim3-5\times$ more overall mass in solids than the MMSN \cp{Chiang2013}.

Proto-planets grow their rocky cores by accreting solid material from within a feeding zone that is proportional to their Hill radius $r_{\mathrm{H}} = a (M_{\mathrm{p}}/3M_{\mathrm{s}})^{(1/3)}$. Therefore integrating $r_{\mathrm{H}}\times \sigma_{\mathrm{solid}}$ over the disk surface area should give us a good idea of the typical maximum mass for rocky planets that form by collisional growth as a function of semi-major axis. Integrating the \ct{Chiang2013} profile in Equation (\ref{mmeneq}), this would predict that $M_{\mathrm{p,max}} \propto a^{0.6} M_{\mathrm{s}}^{-1/2}$ or using the \ct{Zeng2016} mass-radius relation for rocky planets that $R_{\mathrm{trans}} \propto a^{0.16}  M_{\mathrm{s}}^{-0.14}$.

Indeed, this sort of simple calculation is backed up by the results of detailed N-body simulations. For example, \ct{Hansen2013} performed simulated 100 planetary systems using their Monte Carlo N-body model, assuming a $\sigma \propto a^{-3/2}$ density profile. They found a mass distribution with a large amount of scatter but a general trend of increasing mass with increasing semi-major axis. Specifically, they found that the distribution of planet masses in their simulations is well fit by a Rayleigh distribution $f(m) = (m/\sigma^{2}_{\mathrm{m}})e^{-0.5(m/\sigma_{\mathrm{m}})^2}$, with a dispersion $\sigma_{\mathrm{m}} = 7 \, M_{\mathrm{\oplus}}(a/\mathrm{AU})^{0.6}$, the same scaling relation we derived above.

At fixed stellar type then, this would predict that the transition radius for planets that are born rocky should scale as 

\begin{equation}\label{rockytranseq}
R_{\mathrm{trans,bornrocky}} \propto F_{\mathrm{p}}^{-0.08}. 
\end{equation}

In terms of orbital period this is equivalent to  

\begin{equation}\label{rockytranseq2}
R_{\mathrm{trans,bornrocky}} \propto P^{0.11},
\end{equation}

which is in the opposite direction of the trend we predicted for a stripped rocky planet population in Section \ref{evapsec}. 

In terms of a dependence on stellar properties, naively, we would expect $\sigma_{\mathrm{solid}}$ to scale as $\propto M_{\mathrm{disk}} Z_{\mathrm{disk}} \propto M_{\mathrm{s}} Z_{\mathrm{s}}$ \cp[e.g.][]{Kokubo2006}, accounting for the size of the Hill sphere this would predict that $M_{\mathrm{p,max}} \propto M_{\mathrm{s}}^{(1/2)} Z_{\mathrm{s}}$ and therefore that $R_{\mathrm{trans,bornrocky}}$ should increase with stellar mass as $\propto M_{\mathrm{s}}^{0.14}$, again opposite to the predictions of photo-evaporation. However, there is currently debate as to whether there is any evidence for a correlation between planet mass and disk mass or metallicity for non-giant planets \cp[e.g.,][]{Schlaufman2011, Mann2013, Wang2015, Schlaufman2015}, which may be complicated by dispersion in the $Z_{\mathrm{disk}}-Z_{\mathrm{star}}$ relation \cp{Liu2016}. Nonetheless, it seems clear that we would not expect the maximum size of rocky planets to decrease with increasing stellar mass in this scenario, unlike our predictions for a photo-evaporated population, presented in Section \ref{evapsec}.

\begin{figure}
  \begin{center}
    \includegraphics[width=3.5in,height=2.5in]{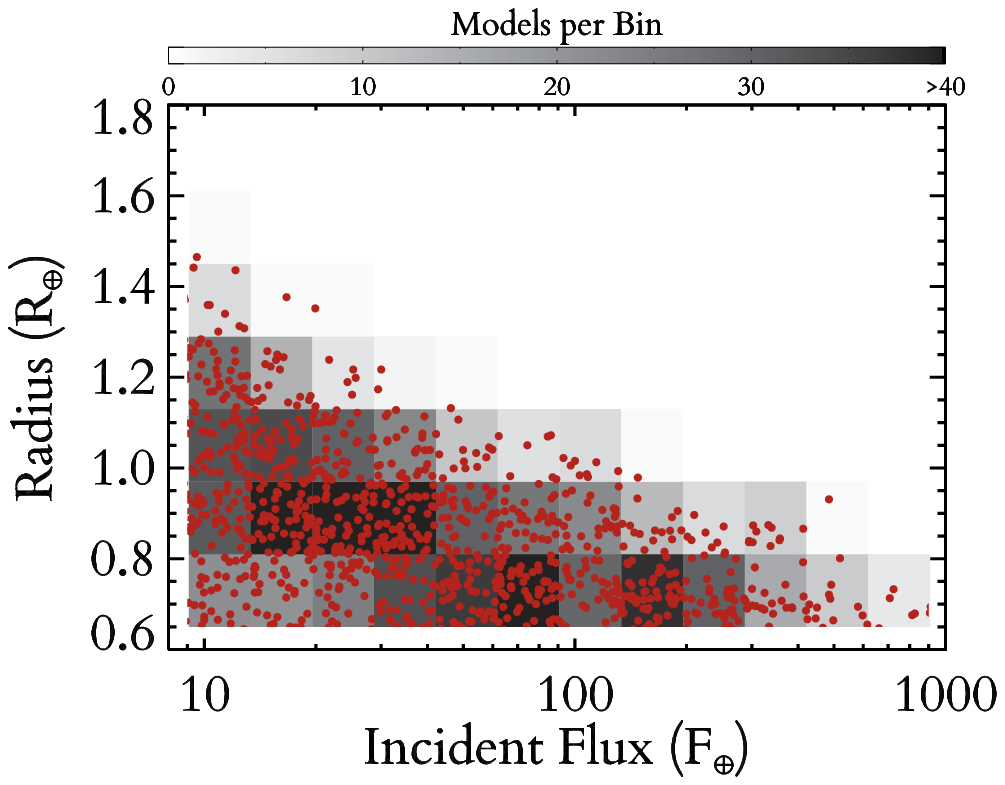}
  \end{center}
  \caption{Similar to Figure \ref{evapfig}, except here we show the prediction from N-body simulations for planets that form without initial envelopes. Unlike in Figure \ref{evapfig}, here we find a maximum size of bare-rocky planets that {\it decreases} with increasing incident flux (Decreasing period.). \label{nbodyfig}}
\end{figure}

To illustrate our predictions for a primordial rocky planet population, we carried out our own large suite of N-body planetary growth simulations. Beginning with the solid surface density profile from equation (\ref{mmeneq}) \cp{Chiang2013}, we consider a disc that extends from $a_{\mathrm{in}} = 0.03$ AU to $a_{\mathrm{out}} = 0.5$ AU and randomly select $\mathcal{F}_{\mathrm{disc}}$ between 0.1 and 1. We first generate our planetary embryos by selecting a random position near the inner edge of the disc and assuming that a planetary embryo grows from this material and that it accretes all of the mass within a feeding zone that is assumed to be 8 Hill radii wide. The next object is then assumed to form at a location
such that it can also accrete all the material within its feeding zone, but without overlapping the feeding zone of the first.  We then progress through the disc turning all the mass in the disc into planetary embryos, ultimatelly forming between about 20 and 40 proto-planetary bodies.


It is then assumed that these planetary bodies are initially all located within the radial extent of the initial disc (i.e., within $a_{\mathrm{out}} \sim 0.5$ AU), that they all have initial eccentricities of $e = 0$, and that they all lie in the same plane. We then carry out N-body integrations, using the hybrid symplectic integrator mercury6 \cp{Chambers1998} and evolve each simulation for 27 Myr, following the approach used by \ct{Dawson2016}. The simulation time step was set to 0.5 days and the integrator was switched from the symplectic integrator to the Burlisch-Stoer integrator if two bodies, $M_{\mathrm{p,1}}$ and M$_{\mathrm{p,2}}$, have a close encounter that brings them within 1 $R_{\mathrm{H}}$, where $R_{\mathrm{H}}$ is the mutual Hill radius

\begin{equation}
R_H = \frac{a_1 + a_2}{2} \left( \frac{ M_{\mathrm{p,1}} + M_{\mathrm{p,2}} } {3 M_{\mathrm{s}}} \right)^{1/3}.
\label{eq:mutual}
\end{equation} 

If two planetary bodies collide then we assume perfect accretion with no fragmentation. However, our general results should be insensitive to this assumption. \ct{Kokubo2010} showed that using more realistic accretion conditions, including fragmentation and hit-and-run collisions, barely affects the mass and number of planets, or even the formation timescale, produced by N-body simulations. In total, we carried out 200 N-body simulations and, after each had been evolved for 27 Myr, were left with 2052 planetary bodies. The results of these simulations are summarized in Figure \ref{nbodyfig}. As expected, we find a maximum radius for the rocky planets formed by collisions that decreases with incident flux, consistent with our analytic predictions and the results of \ct{Hansen2013}, and in stark contrast to the results predicted by our photo-evaporation models in Section \ref{evapsec}. 

Before we move on, however, there is one more important caveat we need to address. In our photo-evaporation calculations in Section \ref{evapsec} we use log uniform core-mass and incident flux distributions which were independent of each other. Yet we have just shown in Section \ref{rockysec} that n-body calculations would predict that on average there should be a correlation between a planets orbit and the typical core mass that it can reach. Our justification for this apparent contradiction is that while the MMEN found by \ct{Chiang2013} represents the typical mass of an extrasolar nebula, scatter in extrasolar disk masses appears to be very large \cp{Chiang2013,Gaidos2017}. As long as the scatter solid mass in extrasolar disks is large enough to provide a large range of core masses at all orbital periods, then in our evaporation scenario, the effects of photo-evaporation should dominate over the increase in average core mass with period and we should find a transition radius that scales as described in equations (\ref{evaptranseq}) and (\ref{evaptranseq2}). This assumption is supported by the significant numbers of ultra-short-period rocky planets with radii $\approx$1.5 $R_{\mathrm{\oplus}}$ \cp{Sanchis-Ojeda2014, Lundkvist2016}. In our born rocky scenario meanwhile, the transition radius, which again is defined as the radius at which the exoplanet planet population switches from a majority rocky to a majority non-rocky population, would reflect the average minimum mass extrasolar nebula, but with significant scatter above and below the transition. If however, the typical scatter in disk masses is small, then the correlations with orbital period or incident flux described in equations (\ref{evaptranseq})/(\ref{evaptranseq2}) and (\ref{rockytranseq})/(\ref{rockytranseq2}) would almost cancel out and we would only be left with a weak correlation that scales as $R_{\mathrm{trans}} \propto F_{\mathrm{p}}^{0.03}.$ or $\propto P^{-0.04}$.

\section{Discussion}

\subsection{Observational Tests}

The results from Sections \ref{evapsec} and \ref{rockysec} present a clear test for exoplanet observers. A clear way to distinguish between a primordial rocky planet population and one that originated as the photo-evaporated cores of sub-Neptunes is to obtain precise mass measurements, particularly with radial velocities, for planets near the observed 1.5 $R_{\mathrm{\oplus}}$ transition for planets receiving a wide range of orbital periods and stellar types. If most of these planets born-rocky like the Earth then we would expect the radius of the rocky to non-rocky transition to {\it increase} with orbital period as $\propto P^{0.11}$, while if they are instead predominantly the evaporated cores of sub-Neptunes then would expect the rocky/non-rocky transition to {\it decrease} with orbital period as $\propto P^{-0.15}$.

Such an observational test should soon be feasible. In early 2018, NASA will launch the Transiting Exoplanet Survey Satellite (TESS) Mission, a two year all sky transit survey of $>$200,000 stars with visual magnitudes of 4-13 \cp{Ricker2014}. Likewise, next year ESA plans to launch the CHaracterising ExOPlanets Satellite (CHEOPS) Mission \cp{Broeg2013}, which will search for transits around bright stars with known radial velocity planets. Together, these new surveys should yield a large new sample of planets around bright nearby stars that can more easily be followed up with radial velocity observations to determine planet masses and therefore constrain planet compositions. 

Moreover, a characterization of the period dependence of the transition should be achievable with a reasonable sample size. For comparison, the current constraints on the rocky/non-rocky transition by \ct{Rogers2015} used the sample of forty-two transiting planets with RV follow-up described by \ct{Marcy2014}. While, this sample is quite large, the majority of these targets provide little information about the transition radius. Only fourteen of the planets in the \ct{Marcy2014} catalog are between 1 and 2  $R_{\mathrm{\oplus}}$, and six of those have uncertainties too large to provide meaningful constraints, i.e., these six have only mass upper limits and those limits are not low enough to test whether those planets have rocky compositions. Therefore, it is only a small subset of $\sim$8 planets that are providing most of our current information on the transition. Accordingly, $\sim$20-30 planets with radii near the current 1.5 $R_{\mathrm{\oplus}}$ transition and 2-3$\sigma$ mass detections, or upper limits that are deep enough to rule out a rocky composition, should be sufficient to test the predictions made here.

Of particular interest will be planets that receive either very high or relatively low levels of irradiation, e.g., $\gtrsim500$  $F_{\mathrm{\oplus}}$ or $\lesssim50$ $F_{\mathrm{\oplus}}$, which for Sun-like stars corresponds to orbital periods $\lesssim3.5$ days or $\gtrsim20$ days respectively. Such a sample should be possible once TESS launches. Using the mock catalog of simulated TESS detections from \ct{Sullivan2015}, we estimate that TESS should find $\sim30$ planets with radii $1.2\, R_{\mathrm{\oplus}}-1.8\, R_{\mathrm{\oplus}} $ around stars with V-band magnitudes brighter than 10 and receiving $>500$  $F_{\mathrm{\oplus}}$, and $\sim$170 planets receiving $<50$  $F_{\mathrm{\oplus}}$. Likewise, planets around later spectral types will also provide a valuable test of the predictions made here, and using the \ct{Sullivan2015} mock catalog, we estimate that TESS should find $\sim80$ planets in this size range around mid-M dwarfs with $T_{eff}<3400$ K and V-mag$<$10.

\subsection{Implications for $\eta_{\mathrm{\oplus}}$}\label{etasec}

One of the primary goals of recent transit surveys, including NASA's {\it Kepler} Mission has been to determine $\eta_{\mathrm{\oplus}}$, the frequency of Earth-sized rocky planets in the habitable zones (HZ) of Sun-like stars. Unfortunately, current surveys are still highly incomplete for planets $\lesssim$1.5 $R_{\mathrm{\oplus}}$ with orbital periods $\gtrsim$200 days \cp[e.g.,][]{Petigura2013a, Foreman-Mackey2014, Burke2015}. Although {\it Kepler} has found several $\gtrsim$1.5 $R_{\mathrm{\oplus}}$ planets in the HZ of sun-like stars \cp{Petigura2013a, Foreman-Mackey2014, Burke2015}, these planets may be gas-rich sub-Neptunes \cp{Rogers2015}. So far only one likely rocky planet Kepler-452b has been found that is potentially in the habitable zone of a Sun-like star \cp{Jenkins2015}. Meanwhile, the shorter period rocky planets, whose occurrence rate is well constrained \cp[e.g.,][]{Fressin2013, Petigura2013a, Foreman-Mackey2014, Burke2015}, may be evaporated cores as described above in Section \ref{evapsec}.  Due to their greater detectability, many recent efforts to estimate the frequency of potentially habitable planets have focused on K and M dwarfs \cp[e.g.,][]{Dressing2013, Kopparapu2013, Morton2014, Dressing2015b, Gaidos2016}, where {\it Kepler} has found many more potentially rocky planets in or near the habitable zone. However, many rocky planets in the habitable zones of late-type stars could also be the products of photo-evaporation, as their stars higher XUV activity means that they could lose initial H/He envelopes of up to $\sim$1\% of their mass \cp{Owen2016}.

All of this means that any efforts to constrain the frequency of true Earth-analogs by looking at the frequency of planets slightly larger and more irradiated than Earth \cp[e.g.,][]{Petigura2013b,Silburt2015}, or around later stellar types, may significantly over-estimate $\eta_{\mathrm{\oplus}}$ by including planets that either are not rocky or have only become rocky due to photo-evaporation. Unfortunately, this is a problem for studies like \ct{Traub2012} that assume the radius and period distributions can be described by separate uncorrelated power-laws, studies like \ct{Burke2015} that use more complex separable broken power-laws, or indeed any study which assumes that planetary radii and irradiation/period distributions are uncorrelated and separable. For this reason it is important to try and determine the origin of the irradiated rocky planet population to understand the significance of this bias. Some methods such as the Gaussian process model used by \ct{Foreman-Mackey2014} or the Approximate Bayesian Computation framework presented by \ct{Hsu2018} allow for the possible of complex non-monotonic correlations between the planetary radii and period distributions, however, thus far these methods have not been applied to the final DR25 {\it Kepler} sample \cp{Thompson2018}.

The need for updated planet occurrence rates is particularly pressing at the moment given current efforts to design future missions and plan observations to search for biomarkers on potentially Earth-like exoplanets. For example, there is currently a large effort to find Earth-sized rocky planets transiting nearby M dwarfs, so that their atmospheres can be characterized by JWST with transmission spectroscopy to look for oxygen and methane in their atmospheres \cp[e.g.,][]{Deming2009, Berta2013, Cowan2015, Barstow2016, Greene2016}. Meanwhile, NASA is currently studying mission concepts for next generation direct imaging missions capable of observing Earth-like planets, including Large UV/Optical/IR Surveyor (LUVOIR) \cp{Crooke2016} and the Habitable Exoplanet Imaging Mission (Habex) \cp{Mennesson2016}, in preparation for the 2020 decadal survey.  Obtaining an accurate estimate for $\eta_{\mathrm{\oplus}}$, which accounts for complex correlations between the radii and period distributions, and an understanding of how this is affected by photo-evaporation will be critical to the success of these efforts.

\section{Summary}

Using models of planet evolution with atmospheric photo-evaporation and Monte Carlo simulations of rocky planet growth we have examined two possible scenarios for the origin of the highly irradiated rocky planet population recently found by NASA's {\it Kepler} Mission. Specifically, we considered the possibility that this rocky planet population is formed from the rocky cores of gas-rich planets that have had their gaseous envelopes stripped by photo-evaporation or impact erosion. We then compared this to a population that initially formed rocky and examined what each of these scenarios predicts for the period dependence of the observed transition between rocky and non-rocky planets. The key points of this study are summarized below.

\begin{itemize}

\item Recent studies have identified a new population of highly irradiated rocky planets with a transition to non-rocky volatile rich sub-Neptunes that occurs around $\sim1.5$ $R_{\mathrm{\oplus}}$.

\item If this rocky planet population originated as the evaporated cores of gas-rich sub-Neptunes, then this transition radius should decrease and rocky planets should become less common at longer orbital periods.

\item If on the other-hand, these planets formed rocky after their disks had already dissipated, then the transition radius should increase with orbital period.

\item With the upcoming launch of new transit missions like TESS and CHEOPS, it should be soon be possible to determine the period dependence of the transition radius with radial velocity follow-up.

\item Of particular interest will be planets that receive either very high or relatively low levels of irradiation, e.g., $\gtrsim500$  $F_{\mathrm{\oplus}}$ or $\lesssim50$ $F_{\mathrm{\oplus}}$, which for Sun-like stars corresponds to orbital periods $\lesssim3.5$days or $\gtrsim20$ days respectively.

\item The difference between these two scenarios has important implications for current efforts to measure $\eta_{\mathrm{\oplus}}$, and will therefore be essential for future missions to study Earth-like planets.

\item In particular, these results show that it is important that studies of planet occurrence rates account for the possibility of complex non-monotonic correlations between planetary radius and period distributions and do not assume that these distributions are seperable.

\end{itemize}

 In reality of course, the two scenarios considered here represent two end member possibilities and the real hot rocky exoplanet population is likely to be a mix of both that were born rocky and of stripped sub-Neptunes. However, since understanding the the relative importance of these two formation channels will be critical to our efforts to search for Earth-like planets, it is important to understand the statistical predictions of these mechanisms so that we can estimate their relative contributions to the observed population.

\section*{Acknowledgements}
We would like to thank the HARPS-N team, Jonathan Fortney, Laura Kreidberg, George McDonald, Kevin Schlaufman, and Angie Wolfgang for their input and many helpful discussions. Eric Lopez is thankful for support from GSFC Sellers Exoplanet Environments Collaboration (SEEC), which is funded by the NASA Planetary Science Division's Internal Scientist Funding Model. The research leading to these results also received funding from the European Union Seventh Framework Programme (FP7/2007-2013) under grant agreement number 313014 (ETAEARTH).

\bsp	
\label{lastpage}
\end{document}